\documentclass[pra,reprint]{revtex4-2} 
\pdfoutput=1

\usepackage{physics}
\usepackage{amsmath}  
\usepackage{amsfonts} 
\usepackage{graphicx} 
\usepackage{float}
\usepackage{epstopdf}
\usepackage{ragged2e}
\usepackage{hyperref}

\begin{document}


\title{Wigner Time Delay in Photoionization: A 1D Model Study}

\author{Karim I. Elghazawy}
\email{kelghaza@purdue.edu} 
\affiliation{Department of Physics and Astronomy, Purdue University, West Lafayette, Indiana 47907 USA}

\author{Chris H. Greene}
\email{chgreene@purdue.edu} 
\affiliation{Department of Physics and Astronomy, Purdue University, West Lafayette, Indiana 47907 USA}
\affiliation{Purdue Quantum Science and Engineering Institute, Purdue University, West Lafayette, Indiana 47907 USA}



\date{\today}

\begin{abstract}
In scattering theory, the Wigner-Smith time delay, calculated through a phaseshift derivative or its multichannel generalization, has been demonstrated to measure the amount of delay or advance experienced by colliding particles during their interaction with the scattering potential. Fetic, Becker, and Milosevic argue that this concept cannot be extended to include photoionization, viewed as a half-scattering experiment. Their argument is based on the lack of information about scattering phaseshifts in the part of the wavefunction (satisfying the ingoing-wave boundary condition) going to the detector. This article aims to test this claim by examining a photodetachment process in a simple 1D model with a short-range symmetrical potential. Using time-dependent perturbation theory with a dipole interaction, the relevant wavepacket of the outgoing particle is analyzed and compared to the free wavepacket as a reference. Our findings confirm that, indeed, a time delay arises in the liberated fragmentation wavepacket, which is expressed as an energy derivative of the scattering phaseshift. Our study highlights that the source of the phaseshift content in the wavepacket arriving at the detector is the dipole matrix element, which is a direct consequence of imposing the ingoing-wave boundary condition. We illustrate our results through numerical simulations of both the non-free and free wavepackets. The amount of the observed time delay is found to be half of that appearing in a typical scattering experiment.
\end{abstract}

\maketitle 

\section{Introduction} 
Photoionization and photodetachment, the processes of electron emission from an atom or negative ion after the absorption of a photon, are core phenomena in the interaction of light with matter, which has played a fundamental role in the development of quantum mechanics. For over a century after Einstein's explanation of the photoelectric effect \cite{Einstein}, photoionization had been assumed to occur instantaneously (i.e. the photoelectron is born instantly upon photon absorption). It was only through recent advances in ultrafast measurement methods that this attosecond process could be temporally resolved. In 2010, Schultze \textit{et al.} \cite{Schultze} reported a finite time delay of the $2p$ photoelectron emission with respect to the $2s$ photoelectron in neon using the attosecond streaking technique. This was followed by a measurement of the relative time delay between $3p$ and $3s$ subshells ionizations of argon in which Kl\"under \textit{et al.} \cite{Argon} employed an interferometry technique called RABBIT (reconstruction of attosecond beating by interference of two-photon transitions). This intensely triggered several studies to apply contemporary theoretical approaches such as the time-dependent \textit{R}-matrix method (RMT) and the random-phase approximation (RPA) in order to explain these time delay measurements \cite{Kheifets_theory1,Moore_theory2,Diagram_theory3,Nagele_theory4,Kheifets_theory5,Feist_theory6,Kheifets_theory7}. However, discrepancies between theory and experiment remained until they were resolved by refined experimental measurements in 2017 \cite{Isinger_exp1} and 2021 \cite{Alex_exp2} as explained in Ref.~\onlinecite{Kheifets_2023}.

The concept of time delay in quantum scattering was originally formulated by Wigner \cite{wigner} and Eisenbud \cite{Eisenbud_thesis} in the context of collision theory. Later, Smith \cite{Smith} further developed the theory and provided a Hermitian operator to quantify the time delay (calculated through the energy derivative of the scattering matrix) experienced during a collision due to the interaction potential. Formally, a quantum scattering event is a continuum-continuum transition from the state of an incoming plane wave to a state of outgoing spherical waves. The interaction potential causes the eigenstate wavefunction to acquire a phaseshift (or more generally, a scattering matrix) that results in a delay (or an advance, as a negative delay) in the peak position of a wavepacket built out of the phase-shifted waves. de Carvalho and Nussenzveig \cite{decarvalho_review} have extensively reviewed the Eisenbud, Wigner and Smith (EWS) time delay theory and related concepts, in which they focused on s-wave scattering. Note that the usual EWS collisional time delay is a function of the collision energy $E$ and can be expressed simply in terms of the energy derivatives of the eigenphaseshifts $\delta_\beta(E)$ of the scattering $S$-matrix, i.e. as:
\begin{equation}
    Q(E) = \sum_\beta 2 \hbar \frac{d\delta_\beta(E)}{dE}.
\end{equation}

Quantum photoionization is a transition from an initial bound state to a final continuum state satisfying the incoming-wave boundary condition \cite{BreitBethe}. The incoming-wave solution is required for photoionization since its component in the wavepacket asymptotically approaches a plane wave wavepacket at long positive times ($t \to \infty$) representing a fragmentation event detected in a definite direction \cite{1982Starace,FanoRau}. Since only the final state involves a scattering eigenstate, photoionization can be regarded, in some sense, as a ``half'' scattering process. The EWS time delay formalism has been extended to include photoionization and photodetachment by Deshmukh \textit{et al.} \cite{Deshmukh1,Deshmukh2} in a treatment that uses the time-reversal conjugate symmetry between incoming- and outgoing-wave solutions. Apparently contesting that point of view, a recent preprint by Fetić, Becker, and Milosević \cite{preprint2022} asserts that the Wigner time delay cannot be inferred from the liberated particle wavefunction, through its phaseshift, and hence cannot be measured in a photoionization setup. Their claim relies primarily on the fact that, in photoionization, the wavepacket of the emitted particle is composed of incoming-wave eigenstates of which the part arriving at the detector (the outgoing free-particle or Coulomb-modified plane wave) carries no information about the scattering phases.

The goal of the present article is to show that although the detected part of the scattering eigenstate does not contribute phaseshift content to the wave front at $r \to \infty$, it is multiplied by the dipole transition matrix element which does contain that scattering phase information. As is illustrated in Section~\ref{theoryA}, this is achieved by analyzing the first order perturbed wavefunction associated with dipole photodetachment of a particle bound in a one-dimensional symmetric potential. It is demonstrated in Section~\ref{theoryB} that, in order to satisfy the incoming-wave boundary condition, real parity eigenstates are linearly combined with complex coefficients involving the scattering eigenphaseshifts, which immediately leads to a phaseshift-dependent dipole transition amplitude. In Section~\ref{calcA}, we calculate the time-dependent wavepacket for a narrow energy bandwidth and track its time delay information. Then, Section~\ref{calcB} analyzes the asymptotic behavior of the time-dependent mean position $\langle x \rangle$ for the photofragmentation wavepacket far away from the short-range potential. Section~\ref{results} presents numerical examples of the wavepacket time propagation with different energy bandwidths. Our examples differ from the majority of current generation time delay experiments, in that we concentrate on systems where final state resonances are predominant. Finally, we conclude in Section~\ref{conclusion} and offer further implications of our findings. Much of our analysis is inspired by a recent review of this subject by Rau \cite{Rau_2021}.

\section{Theory} \label{theory}
    \subsection{First Order Perturbation Theory} \label{theoryA}
Atomic units ($\hbar=m=e=1$) are used throughout the paper. Consider the unperturbed one-dimensional Hamiltonian\begin{equation}
    \hat{H}_0=-\frac{1}{2}\frac{d^2}{dx^2}+V_0(x),
    \label{ham}
\end{equation}
where $V_0(x)$ is a localized potential symmetric in $x$ (i.e. $V_0(-x)=V_0(x)$), and has only one bound state $\ket{\psi_g}$ and a continuum of energy-normalized scattering states $\ket{\psi_E}$. These states satisfy
\begin{equation}
\begin{aligned}
       \hat{H}_0  \ket{\psi_g} &= \epsilon_g  \ket{\psi_g}, &\epsilon_g < 0  \\  \hat{H}_0\ket{\psi_E} &= E \ket{\psi_E},     &E>0. 
\end{aligned}
\label{eigen}
\end{equation}
If now a time-dependent perturbation $\hat{V}(t)$ is introduced, the full Hamiltonian becomes $  \hat{H}=\hat{H_0}+\hat{V}(t)$. Since a bound to continuum transition is of interest here, the system is assumed to be initially in the state $\ket{\psi_g}$, (i.e.  $\ket{\psi(t_0)}=\ket{\psi_g}$). The time-dependent Schr\"odinger solution can be written as:
\begin{equation}
        \ket{\psi(t)} = C_g(t)  \ket{\psi_g} e^{- i \epsilon_g t} + \sum_{\lambda} \int C^{\lambda}_E(t)  \ket{\psi^{\lambda}_E} e^{- i E t}\, dE,
    \label{ansatz}
\end{equation}
  with $C_E(t_0)=0, \,C_g(t_0)=1$, where the index $\lambda$ labels the degenerate eigenstates at energy $E$. By exploiting the eigenstate orthogonality and Eqs.~(\ref{eigen}), the perturbative first order wavefunction correction is obtained:
\begin{equation}
    C^{\lambda}_E(t) = -i \int_{t_0}^t e^{i(E-\epsilon_g)t'} \bra{\psi^{\lambda}_E}\hat{V}(t')\ket{\psi_g} dt'.
    \label{correction}
\end{equation}

Now specialize to a Gaussian light pulse, i.e.
\begin{equation}
    \hat{V}(t) = \hat{d}\, e^{- i w_0 t}\, e^{-\frac{t^2}{2s^2}},
\end{equation}
where $\hat{d}$ is the dipole operator. After substituting this perturbation into Eq.~(\ref{correction}) and evaluating the time integral from $t_0=-\infty$, the coefficient equals
\begin{align}
    C^{\lambda}_E(t) &= -i\bra{\psi^{\lambda}_E}\hat{d}\ket{\psi_g} \int_{-\infty}^t e^{i(E-\epsilon_g-w_0)t'} e^{-\frac{t'^2}{2s^2}} dt' \nonumber \\
    &= -i\sqrt{\frac{\pi}2}s \bra{\psi^{\lambda}_E}\hat{d}\ket{\psi_g}  e^{-\frac{s^2}{2}(E-E_0)^2} \nonumber \\
    &\times \left(1+ \erf\left[\frac{t}{\sqrt{2}s}-i\frac{(E-E_0)s}{\sqrt{2}}\right]\right),      \label{corrV}
\end{align}
where $E_0 = \epsilon_g \,+\, w_0$ and $\erf$ is the error function. Inserting Eq.~(\ref{corrV}) in Eq.~(\ref{ansatz}) gives the wavefunction up to first order: $\ket{\psi(t)} = \ket{\psi_g} e^{- i \epsilon_g t} + \ket{\psi_{p}(t)}$, where
\begin{align}
    \ket{\psi_{p}(t)}&= -i\sqrt{\frac{\pi}2}s \sum_{\lambda}\int  \bra{\psi^{\lambda}_E}\hat{d}\ket{\psi_g}e^{-\frac{s^2}{2}(E-E_0)^2}  \ket{\psi^{\lambda}_E} \, \nonumber\\ 
    &\times e^{- i E t} \left(1+ \erf\left[\frac{t}{\sqrt{2}s}-i\frac{(E-E_0)s}{\sqrt{2}}\right]\right)dE 
   \label{wavepacket}
   \end{align}
is the part of the wave function relevant to photoionization/photodetachment. Note how the factor in round parentheses goes to zero for $t \to - \infty$ (since $\erf[-\infty] = 0$) which conveniently satisfies the initial condition.
\subsection{Defining The Potential} \label{theoryB}
\begin{figure}[ht!]
    \centering
    \includegraphics[width=3.375in]{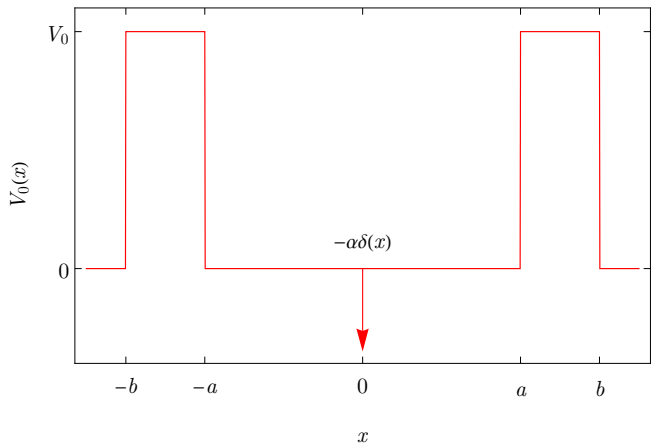}
    \caption{
    The potential energy $V_0(x)$ in Eq.~(\ref{potential_eq}) is plotted versus $x$. The Dirac-delta well is represented by a downward arrow at the origin.}
    \label{pot}
\end{figure}

\begin{figure}[ht!]
    \centering
    \includegraphics[width=3.375in]{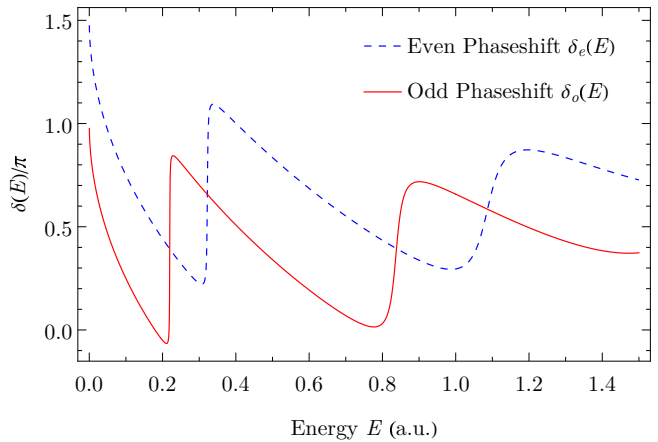}
    \caption{The even- and odd-parity phaseshifts (divided by $\pi$) are plotted versus energy for $a=4, \ b=6, \ V_0=1$ and $\alpha=1$. Each rapid rise of $\delta/\pi$ by approximately unity corresponds to a scattering resonance.}
    \label{phaseshift}
\end{figure}

Consider the short-range potential 
energy \begin{equation}
    V_0(x) = -\alpha \delta(x) + \begin{cases}
    V_0 & a < \abs{x} < b \, , \\
    0 & \abs{x} < a ~ \textrm{or} ~ \abs{x} > b
    \end{cases}
    \label{potential_eq}
\end{equation}
to be the unperturbed potential in Eq.~(\ref{ham}).
Figure~\ref{pot} shows a sketch of this potential energy where the Dirac-delta well is represented by a downward arrow at the origin.
Clearly, this potential admits only one bound state (for $\alpha > 0$, $V_0>0$) which is the purpose of the Dirac-delta function. For a wide enough well (i.e. $a \gg 1/\alpha$), this state can be taken, to a very good approximation, to be the bound state of the Dirac-delta potential alone:
\begin{equation}
    \psi_g(x) 
    \approx {} \sqrt{\alpha} e^{- \alpha \abs{x}}, ~ \epsilon_g = -\frac{\alpha^2}{2}.
\end{equation}
Note that this state is well localized inside the double barrier potential and is negligible outside the well. Since $V_0(x)$ is a symmetric potential, scattering eigenstates can be chosen to have a definite parity. The even and odd energy-normalized eigenstates are
\begin{subequations}
\begin{align}
        \psi_E^e(x) &= \frac1{\sqrt{\pi k}} \label{even state}
        \begin{cases}
            N_1^e \cos( k \abs{x}+\beta) & \abs{x}\leq a \, ,\\
            N_2^e \cos( q x \pm \Delta_e ) & a < \abs{x} \leq b \, ,\\
            \cos( k x \pm \delta_e) &   \abs{x}>b \, ,
        \end{cases}
       \\
        \psi_E^o(x) &= \frac1{\sqrt{\pi k}} \label{odd state}
        \begin{cases}
           N_1^o \sin( k x) & \abs{x} \leq a \, ,\\
           N_2^o \sin( q x \pm \Delta_o ) & a < \abs{x} \leq b \, , \\
           \sin( k x \pm \delta_o) &  \abs{x}>b \, ,
        \end{cases}
\end{align}
\label{parity}
\end{subequations}
respectively, where the plus (minus) sign is for $x>0$ ($x<0$), $k = \sqrt{2E}$ and $q=\sqrt{2(E-V_0)}$. The parameters $N_1^e, N_2^e, N_1^o, N_2^o, \beta, \Delta_e, \Delta_o, \delta_e, \delta_o$ are energy-dependent constants determined by imposing the relevant boundary conditions at $x= 0, \ a, \ b$.
Figure~\ref{phaseshift} shows the even and odd phaseshifts as functions of energy for  $a=4, \ b=6, \ V_0=1$ and $\alpha=1$, where shape resonances are identified through a sudden rise by approximately $\pi$. These resonances correspond to the bound states of the corresponding finite well of width 4 and depth 1 a.u.

For a photoionization process, the energy eigenstates used in the energy expansion in Eq.~(\ref{wavepacket}) should be chosen to satisfy the incoming-wave boundary condition. In one dimension, there are two choices corresponding to different experimental observations: $\ket{\psi_E^R}$ and $\ket{\psi_E^L}$, for which particles are detected on the right or the left, respectively. These energy-normalized solutions have the asymptotic form
\begin{subequations}
\begin{align}
    \psi_E^R(x) &\xrightarrow[x \to \pm \infty]{}  \frac1{\sqrt{2 \pi k}} \big( e^{+i k x} + f^R_\pm (E) \, e^{-i k \abs{x}}\big), \label{psiR(x)}\\
    \psi_E^L(x) &\xrightarrow[x \to \pm \infty]{}  \frac1{\sqrt{2 \pi k}} \big( e^{-i k x} + f^L_\pm (E) \, e^{-i k \abs{x}}\big),
\end{align}
\label{incoming}
\end{subequations} where $f_\pm(E)$ are energy-dependent constants. The incoming-wave states are linear combinations of the real, definite parity states (Eqs.~(\ref{parity})), namely  
\begin{subequations}
    \begin{align}
        \ket{\psi_E^{R}} &=  \frac1{\sqrt{2}} \big(e^{-i \delta_e} \ket{\psi_E^e} + i e^{-i \delta_o}\ket{\psi_E^o}\big) \label{linearcomR}\\
        \ket{\psi_E^{L}} &=\frac1{\sqrt{2}} \big( e^{-i \delta_e} \ket{\psi_E^e} - i e^{-i \delta_o}\ket{\psi_E^o} \big)
\end{align}
\label{linearcombination}
\end{subequations}
This implies (using Eqs.~(\ref{parity}))
\begin{equation} \label{f(E)}
\begin{aligned}
    f^R_+(E) = f^L_-(E) &= \frac12 (e^{-2i\delta_e} - e^{-2i\delta_o}), \\
    f^R_-(E) = f^L_+(E) &= \frac12(e^{-2i\delta_e} + e^{-2i\delta_o})-1. 
\end{aligned}
\end{equation}
Turning to our wave-packet in Eq.~(\ref{wavepacket}) and considering the long time limit (more specifically $t \gg s$), which is appropriate for photofragmentation, one has
\begin{align}
    &\ket{\psi_{p}(t)} \xrightarrow[t \to \infty]{} \nonumber \\
    &-i\sqrt{2\pi}s \int  \bra{\psi_E^R}\hat{d}\ket{\psi_g}e^{-\frac{s^2}{2}(E-E_0)^2}\ket{\psi_E^R} e^{- i E t}\,dE \nonumber \\
&-i\sqrt{2\pi}s \int \bra{\psi_E^L}\hat{d}\ket{\psi_g}e^{-\frac{s^2}{2}(E-E_0)^2} \ket{\psi_E^L} e^{- i E t}\,dE \nonumber 
\\ &\phantom{\ket{\psi'(t)}} \equiv \ket{\psi_R(t)} + \ket{\psi_L(t)}
\label{psiR&L}
\end{align}
where the scattering continuum is divided into the incoming-wave states $\ket{\psi_E^R}$ and $\ket{\psi_E^L}$. 

\section{Calculations} \label{calc}
\subsection{The Simple Case} \label{calcA}
This subsection analyzes the time delay information contained in the photodetachment wavepacket $\ket{\psi_p(t)}$ in the case of a narrow energy bandwidth (large $s$) focusing only on the right-escaping wavepacket $\ket{\psi_R(t)}$. Consider, for definiteness, the first integral of Eq.~(\ref{psiR&L}) for the eigenstates $\ket{\psi_E^R}$. This corresponds to the detector being placed at $x \to +\infty$, to observe only a particle escaping to the right. Now, from Eq.~(\ref{linearcomR}), 
\begin{equation}
\bra{\psi_E^R}\hat{d}\ket{\psi_g} =- \frac{i e^{i \delta_o}}{\sqrt{2}}\bra{\psi_E^o}\hat{d}\ket{\psi_g},
    \label{dipole}
\end{equation}
where the dipole transition amplitude to $\ket{\psi_E^e}$ vanishes, since $\ket{\psi_g}$ has even parity. Insertion of Eq.~(\ref{dipole}) into the definition for $\ket{\psi_R(t)}$ gives
\begin{align} 
\ket{\psi_R(t)} = -\sqrt{\pi}s &\int \bra{\psi_E^o}\hat{d}\ket{\psi_g}e^{-\frac{s^2}{2}(E-E_0)^2} \nonumber \\ &\times \ket{\psi_E^R} e^{- i E t} e^{i \delta_o}\,dE \label{psiR(t)}.
\end{align}
Using Eqs.~(\ref{psiR(x)}) and (\ref{f(E)}) to analyze the asymptotic behavior of the wavepacket far away from the localized potential, it follows that
\begin{align}
   &\psi_R(x,t) 
   \to -s\int  \frac{e^{i \delta_o}}{\sqrt{2k}} \bra{\psi_E^o}\hat{d}\ket{\psi_g}e^{-\frac{s^2}{2}(E-E_0)^2}e^{-i E t} \times \nonumber \\  
    &\begin{cases}
        \frac12 (e^{-2i\delta_e} - e^{-2i\delta_o}) e^{-i k x} + e^{+i k x} & x \to +\infty\\
        \frac12 (e^{-2i\delta_e} + e^{-2i\delta_o}) e^{+i k x} & x \to -\infty
    \end{cases} 
    dE. \label{psiR(t)b}
\end{align}

Now comes the essential step: since the Gaussian envelope $e^{-\frac{s^2}{2}(E-E_0)^2}$ is peaked around $E = E_0$, it is legitimate to Taylor expand all the energy-dependent phases ($\delta_e, \delta_o, k,\,$and $E$) in the integrand around this particular energy. Of course, this is only valid for Gaussian bandwidths (determined by $s$) much narrower than the energy scale over which these phases vary. Stated differently, these energy-dependent phases are assumed to be slowly varying around $E_0$. For example, the phaseshifts $\delta_e,\, \delta_o$ might have resonances close to $E_0$ which lead to sharp peaks in the derivatives $\delta_e'(E),\, \delta_o'(E)$. In such a case near a resonance, the Gaussian wavepacket energy width $\propto 1/s$ must be much smaller than the resonance width(s). Consequently, one can write, as a first approximation, 
\begin{equation}
    \begin{aligned}
        \delta_{e/o}(E) &\approx \delta_{e/o}(E_0) + \frac{d\delta_{e/o}}{dE}\Big|_{E_0} (E-E_0) \\
        &\equiv \delta_{e/o}^0 + \delta'^0_{e/o}(E-E_0),\\
        k(E) &\approx k(E_0) + \frac{dk}{dE}\Big|_{E_0} (E-E_0) \\
        &\equiv k_0 + \frac1{v_0}(E-E_0), \ \textrm{with}\ v_0 = \sqrt{2E_0}, \\
        E &= E_0 + (E-E_0).
    \end{aligned}
    \label{linear expansion}
\end{equation}
As for the energy-dependent dipole matrix element $\bra{\psi_E^o}\hat{d}\ket{\psi_g}$, it is sufficient to approximate it by its value at $E=E_0$ because while resonances in phaseshifts cause sharp peaks in phaseshifts first derivative, they cause sharp peaks in the dipole matrix element itself (as a function of energy) not its derivative.
\begin{figure}[ht!]
\centering
\includegraphics[width=3.375in]{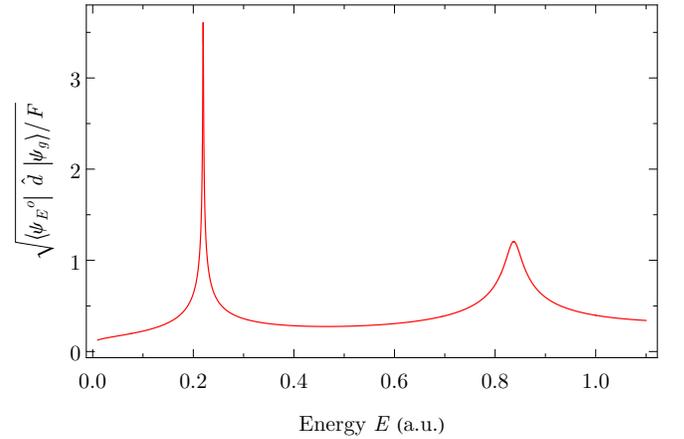}
\caption{The dipole matrix element square root is plotted versus final state energy for $a=4, \ b=6, \ V_0=1$ and $\alpha=1$. Two peaks can be seen at the same resonant energies seen in the odd-parity phaseshift in Fig.~\ref{phaseshift}}
\label{dipole_graph}
\end{figure}
For example, we can calculate the odd dipole transition amplitude $\bra{\psi_E^o}\hat{d}\ket{\psi_g}$ as a function of the final eigenstate energy by writing:
\begin{align}
    \bra{\psi_E^o}\hat{d}\ket{\psi_g} &= F \int \psi_E^o(x) \ x \ \psi_g(x) \, dx  \nonumber \\
    &= \frac{F N_1^o(E) \sqrt{\alpha}}{\sqrt{\pi k}} \int_{-\infty}^{\infty} \sin(k x) \ x \   e^{- \alpha \abs{x}} \, dx \nonumber \\
    &= \frac{F N_1^o(E) \sqrt{\alpha}}{\sqrt{\pi k}} \frac{4 \ \alpha \ k}{(k^2 + \alpha^2)^2},
\end{align}
where $F$ is the electric field amplitude. In the preceding integral, the scattering state $\psi_E^o(x)$ was replaced by its form inside the potential well ($\abs{x} < a$) since the bound state $\psi_g(x)$ is well localized inside the potential well (i.e. $a \gg 1/\alpha$). As can be seen in Fig.~\ref{dipole_graph}, two sharp peaks exist in the odd dipole matrix element $\bra{\psi_E^o}\hat{d}\ket{\psi_g}$ at the same resonant energies observed for the odd-parity phaseshift (see Fig.~\ref{phaseshift}). These peaks are due to the existence of very narrow energy ranges over which the state $\ket{\psi_E^o}$ has a relatively large overlap with $\ket{\psi_g}$, which leads (usually) to a maximized $\bra{\psi_E^o}\hat{d}\ket{\psi_g}$. These resonance states are also referred to as \textit{quasi-bound} states. Thus \begin{equation}
   \frac1{\sqrt{k}} \bra{\psi_E^o}\hat{d}\ket{\psi_g} \approx \frac1{\sqrt{k_0}}\bra{\psi^{o}_{E_0}}\hat{d}\ket{\psi_g}
   \label{dipole_E0}
\end{equation} 

Exploiting Eqs.~(\ref{linear expansion}) and (\ref{dipole_E0}), a sample integral in Eq.~(\ref{psiR(t)b}) (second term in the $x \to \infty$ case) looks like
\begin{align}
I =  &-\frac{s}{\sqrt{2k_0}}\bra{\psi_{E_0}^o}\hat{d}\ket{\psi_g} e^{i (k_0 x -E_0 t + \delta_o^0)} \nonumber\\ &\times \int e^{-\frac{s^2}{2}(E-E_0)^2} e^{i (\frac{x}{v_0}-t+\delta'^0_o)(E-E_0)} \, dE
 \nonumber\\
=&-\sqrt{\frac{\pi}{k_0}}\bra{\psi^{o}_{E_0}}\hat{d}\ket{\psi_g} \nonumber e^{i (k_0 x -E_0 t + \delta_o^0)} \\ &\times e^{-\frac12 \big[\frac{x- v_0 (t-\delta'^0_o)}{s v_0}\big]^2},
\end{align}
where similar forms exist for the remaining integrals in $  \psi_R(x,t)$. Hence, one finds
\begin{align}
    \psi_R(x,t) &\xrightarrow[x \to \pm \infty]{t \to \infty} -\sqrt{\frac{\pi}{k_0}}\bra{\psi^{o}_{E_0}}\hat{d}\ket{\psi_g} e^{-i E_0 t}  \nonumber\\ \times \big[
    &\frac12e^{-i k_0 \abs{x}} e^{i\delta^0_o-2i\delta_e^0}e^{-\frac12 \big[\frac{\abs{x} + v_0 (t+2\delta'^0_e-\delta'^0_o)}{s v_0}\big]^2}  \nonumber\\ \mp \, &\frac12e^{-i k_0 \abs{x}} e^{-i\delta^0_o}e^{-\frac12 \big[\frac{\abs{x} + v_0 (t+\delta'^0_o)}{s v_0}\big]^2} \nonumber
    \\  + \, &e^{i k_0 x} e^{i \delta_o^0} e^{-\frac12 \big[\frac{x- v_0 (t-\delta'^0_o)}{s v_0}\big]^2} (+\infty) \big] 
\end{align}
Evidently, for $t \to \infty$, only the outgoing-wave terms survive, which is an immediate consequence of the incoming-wave boundary condition (Eq.~(\ref{psiR(x)})). Therefore 
\begin{equation}
     \abs{\psi_R(x,t)}^2 \xrightarrow[\abs{x} \to \infty]{t \to \infty} \frac{\pi}{k_0}\abs{\bra{\psi^{o}_{E_0}}\hat{d}\ket{\psi_g}}^2  e^{-\frac12 \big[\frac{x- v_0 (t-\delta'^0_o)}{s v_0}\big]^2},
     \label{result1}
\end{equation}
which is a Gaussian moving to the right with speed $v_0$ and its peak is located at $x =  v_0 (t-\delta'^0_o)$. This suggests that it is time delayed by $\tau_d = \delta'_o(E_0)$ relative to the free wavefunction not encountering the double barrier (i.e. $V_0 = 0 =\delta_o$) which would normally have its peak at $x = v_0 t$.
\subsection{The General Case} \label{calcB}
In the preceding subsection, it was shown that the outgoing photoparticle wavepacket $\ket{\psi_R(t)}$ contains a signature of time delay, as manifested by the position of its peak in the case of a narrow energy bandwidth (large $s$). However, this signature is not always useful since the wavefunction may generally become skewed or even have more than one peak (for a wider energy bandwidth). In this case, one loses the ability to interpret the difference between peaks of the free and non-free wavepackets as a sign of time delay. Consequently, it is more instructive to analyze the motion of the mean position of the right-escaping wavepacket $\langle x\rangle$, as a function of time in order to gain clearer insight into the time delay information contained in the wavepacket $\ket{\psi_R(t)}$. This is the quantity 
\begin{equation}
    \langle x \rangle \equiv \frac{\bra{\psi_R} x \ket{\psi_R} }{\bra{\psi_R} \ket{\psi_R}},
\end{equation}
computed in the long time limit, long enough for the wavepacket to travel completely outside the potential well.

Use of Eq.~(\ref{psiR(t)}) leads to \begin{align}
    \bra{\psi_R} x \ket{\psi_R} = \pi s^2 \int  d(E') g(E') e^{-i(\delta_o(E')- E' t)}\nonumber \\ \times \, d(E) g(E) e^{i(\delta_o(E)- E t)}\bra{\psi_{E'}^R}x\ket{\psi_E^R} \,dE' \,dE, \label{<x>numerator}
\end{align}
where $d(E) \equiv \bra{\psi_E^o}\hat{d}\ket{\psi_g}$ and $g(E) \equiv e^{-\frac{s^2}{2}(E-E_0)^2}$. By exploiting the eigenstate energy-normalization, the following identity can be derived (see appendix \ref{Appendix}):
\begin{equation}
    \bra{\psi_{E'}^R}x\ket{\psi_E^R} \xrightarrow[t \to \infty]{} -i \sqrt{kk'} \frac{d}{dE} \big[\delta(E-E')\big]
    \label{Dirac}
\end{equation}
where $\delta()$ is the Dirac-delta function. The time limit here is understood to be evaluated after putting the scattering state into a time-dependent wavepacket energy integral. Insertion of Eq.(\ref{Dirac}) into Eq.~(\ref{<x>numerator}) and integration by parts gives
\begin{align}
    &\begin{aligned} \nonumber
      &\bra{\psi_R} x \ket{\psi_R} = i\pi s^2 \int \sqrt{k'} \ d(E') g(E') e^{-i(\delta_o(E')- E' t)} \\ &\times \frac{d}{dE} \left[\sqrt{k} \ d(E) g(E) e^{i(\delta_o(E)- E t)}\right] \delta(E-E') \,dE' \,dE 
    \end{aligned} \\
    &\phantom{\bra{\psi_R} x \ket{\psi_R}}= \,i\pi s^2 \int  \sqrt{k} \ d(E) g(E) e^{-i(\delta_o(E)- E t)}\nonumber\\ &\phantom{\bra{\psi_R} x \ket{\psi_R}}\times  \frac{d}{dE} \left[\sqrt{k} \ d(E) g(E) e^{i(\delta_o(E)- E t)}\right] \,dE, 
    \end{align}
where the $E'$ integral has been evaluated in the last step. Expanding the derivative 
\begin{align}
    \bra{\psi_R} x \ket{\psi_R} =  & \ \pi s^2 \int  k (t - \delta'_o) \nonumber [d(E) g(E)]^2 \,dE \  \\ + & \ \frac{i \pi s^2}{2} \int 
    \frac{d}{dE} \left[\sqrt{k} \ d(E) g(E) \right]^2 dE \nonumber\\
    = & \ \pi s^2 \int  k (t - \delta'_o) \left[d(E) g(E)\right]^2 \,dE
\end{align}
where the second integral vanishes for any finite energy bandwidth. Similarly for the denominator
\begin{equation}
    \bra{\psi_R} \ket{\psi_R}
    = \pi s^2 \int   \left[d(E) g(E)\right]^2 dE,
\end{equation}
since $\bra{\psi_{E'}^R}\ket{\psi_E^R}=\delta(E-E') $. The final result is then simply: 
\begin{equation}
    \langle x \rangle = A \, t - B,
\end{equation}
with 
\begin{equation} \label{result2}
\begin{aligned}
    A &= \frac{\int k \left[\bra{\psi_E^o}\hat{d}\ket{\psi_g}e^{-\frac{s^2}{2}(E-E_0)^2}\right]^2 \,dE }{\int \left[\bra{\psi_E^o}\hat{d}\ket{\psi_g}e^{-\frac{s^2}{2}(E-E_0)^2}\right]^2 \,dE }, \\
    B &= \frac{\int k \, \delta'_o \left[\bra{\psi_E^o}\hat{d}\ket{\psi_g}e^{-\frac{s^2}{2}(E-E_0)^2}\right]^2 \,dE }{\int \left[\bra{\psi_E^o}\hat{d}\ket{\psi_g}e^{-\frac{s^2}{2}(E-E_0)^2}\right]^2 \,dE },
\end{aligned}
\end{equation}
demonstrating that $\ket{\psi_R(t)}$ is time-delayed by $\tau_d = B/A$, relative to the free final state wavefunction for which $V_0 = 0 = \delta_o$.

\section{Results and Discussion} \label{results}
Eq.~(\ref{result1}) shows an initial evidence of a delay time contained in $\ket{\psi_R(t)}$ given directly by the odd phaseshift energy derivative. Although the part of the incoming-wave scattering state $\ket{\psi_E^R}$ that survives at $t \to \infty$ when put into the wavepacket integral (i.e. the outgoing-wave part) does not contain any phaseshift information (see Eq.~(\ref{psiR(x)})), the full time-dependent wavepacket $\ket{\psi_R(t)}$ still acquires phaseshift content through the dipole matrix element $\bra{\psi_E^R}\hat{d}\ket{\psi_g}$ multiplying the scattering state $\ket{\psi_E^R}$, more specifically, through its complex phase $e^{i \delta_o}$. This is due to the crucial linear combination in Eq.~(\ref{linearcomR}). Assuredly, the calculations in Section~\ref{calcA} could have been done using the parity eigenstates $ \{ \ket{\psi_E^e}, \ket{\psi_E^o} \}$ in the energy expansion (Eq.~(\ref{psiR&L})) instead of $\{ {\ket{\psi_E^R}, \ket{\psi_E^L}} \}$. In the former basis, the energy expansion for $\ket{\psi_p(t)}$ would have only one integral containing $\bra{\psi_E^o}\hat{d}\ket{\psi_g} \ket{\psi_E^o}$ (the other vanishes by parity). This is equivalent to combining the two integrals in Eq.~(\ref{psiR&L}) since (in light of Eqs.~(\ref{linearcombination}))
$
    \bra{\psi_E^R}\hat{d}\ket{\psi_g}\ket{\psi_E^R} + \bra{\psi_E^L}\hat{d}\ket{\psi_g}\ket{\psi_E^L}= \bra{\psi_E^o}\hat{d}\ket{\psi_g}\frac{-i e^{i \delta_o}}{\sqrt{2}}\left(\ket{\psi_E^R} -\ket{\psi_E^L}\right)
        = \bra{\psi_E^o}\hat{d}\ket{\psi_g} \ket{\psi_E^o}.
$
In this perspective, however, the phaseshift (time delay) content is mainly encoded in the odd-parity scattering state itself (see Eq.~(\ref{odd state})). But irrespective of the chosen basis, the net result is that the wavepacket is time-delayed by $\tau_d = \delta'_o(E_0)$. Note that this is precisely half the time that normally arises in a standard scattering process, consistent with the perception that photofragmentation can be regarded as a half-scattering process.

Whereas the analysis leading to Eq.~(\ref{result1}) is valid only for narrow energy bandwidths (large $s$), the analysis in Section~\ref{calcB} is valid for arbitrary bandwidths. It provides intuition about the wavepacket aggregate motion far away from the potential range, which is generally more complicated than just one peak moving with speed $v_0=\sqrt{2E_0}$ and time-delayed by $\tau_d= \delta'_o(E_0)$. Eqs.~(\ref{result2}) demonstrate that more general expressions, valid for arbitrary bandwidths, for the group velocity and time delay emerges after convolving the relevant quantity with the envelope function $\left[d(E) g(E)\right]^2$, the probability of having the stationary state $\ket{\psi_E^R}$ in the wavepacket mixture (see Eq.~(\ref{psiR(t)})). Stated more simply, Eqs.~(\ref{result2}) represent a weighted-average of $k$ and $k \, \delta'_o$ with $\left[d(E) g(E)\right]^2$ as the weighting function. 

\begin{figure}[ht!]
\centering
\includegraphics[width=3.375in]{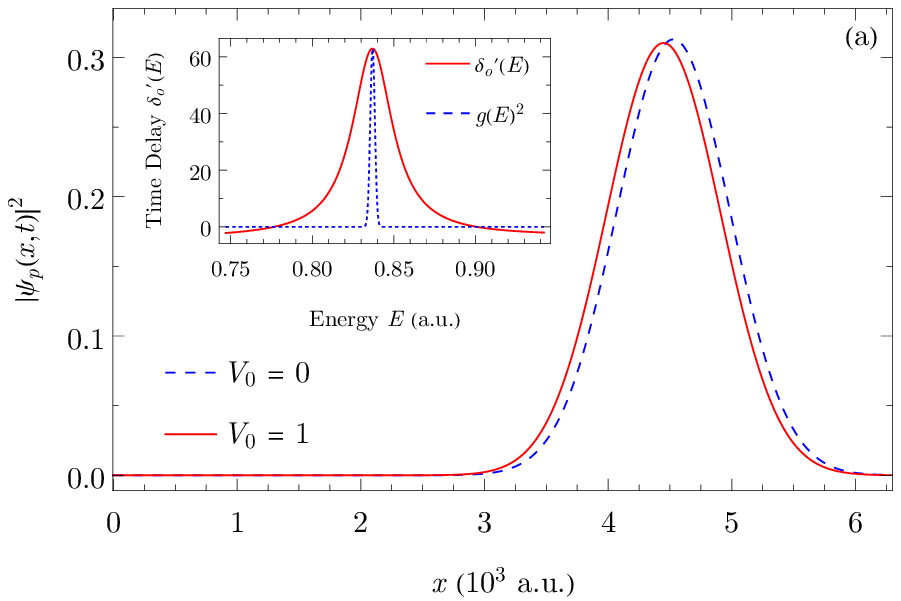}
\\ \vspace{12pt}
\includegraphics[width=3.375in]{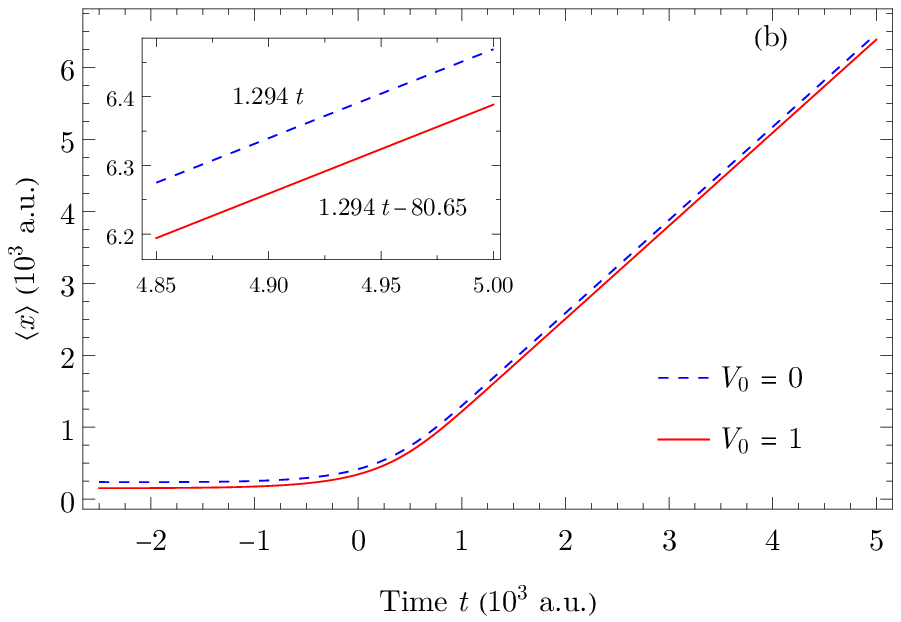}
\caption{(First Case) (a): the squared norm of the wavepackets for $V_0=0$ (blue dashed curve) and $V_0 \neq 0$ (red solid curve) are plotted versus distance at a large positive time. The two wavepackets have been rescaled to have the same normalization. The inset exhibits the Gaussian envelope $g(E)^2$, arbitrarily scaled, (blue dashed curve) over-plotted with the odd phaseshift energy derivative $\delta'_o(E)$ (red solid curve). (b): $\langle x \rangle$ for $V_0=0$ (blue dashed curve) and $V_0 \neq 0$ (red solid curve) is plotted versus time. The inset focuses on the asymptotics and shows their straight line fit equations. The parameters values are: $a=4,\ b=6,\ E_0=0.83695$,\ $s=500$, and $\alpha=1$.}
\label{case1}
\end{figure}
We now present numerical computations of the time propagated wavepacket $\psi_p(x,t)$ in Eq.~(\ref{wavepacket}), carried out for right-escaping scattering states $\ket{\psi_E^R}$ with different values of the parameters $a, b, E_0$ and $s$. In each case, $\langle x \rangle$ has been computed as a function of time, using the propagated wavepacket $\psi_p(x,t)$, and its asymptote was fitted to a straight line to extract the coefficients $A$ and $B$ defined in Eqs.~(\ref{result2}). All calculations were performed for both the non-free and free-particle final state wavepacket (obtained by setting $V_0=0$) for comparison.

Figure~\ref{case1} illustrates the case of a narrow energy bandwidth relative to the width of the resonance in $\delta'_o(E)$, which is the validity domain of Section~\ref{calcA}, as shown in the inset of Fig.~\ref{case1} (a). One can see in (a) that the non-free wavepacket far away from the potential is simply a Gaussian with a delayed peak relative to the free wavepacket peak, in agreement with Eq.~(\ref{result1}). In this case, the space delay information inferred from either the difference of the two Gaussian peaks positions (Fig.~\ref{case1} (a)) or the $B$ coefficient in the straight line fit of the non-free $\langle x \rangle$ curve (Fig.~\ref{case1} (b) inset) is identical. This is simply due to the fact that the average value coincides with the peak position for a symmetric distribution.

\begin{figure}[ht!]
  \centering
  \includegraphics[width=3.375in]{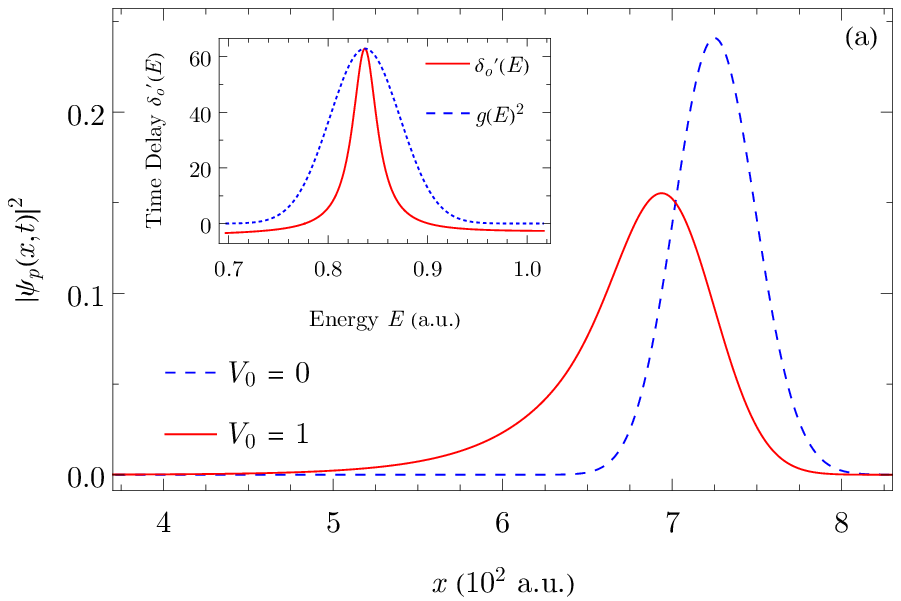}
  \\ \vspace{12pt}
  \includegraphics[width=3.375in]{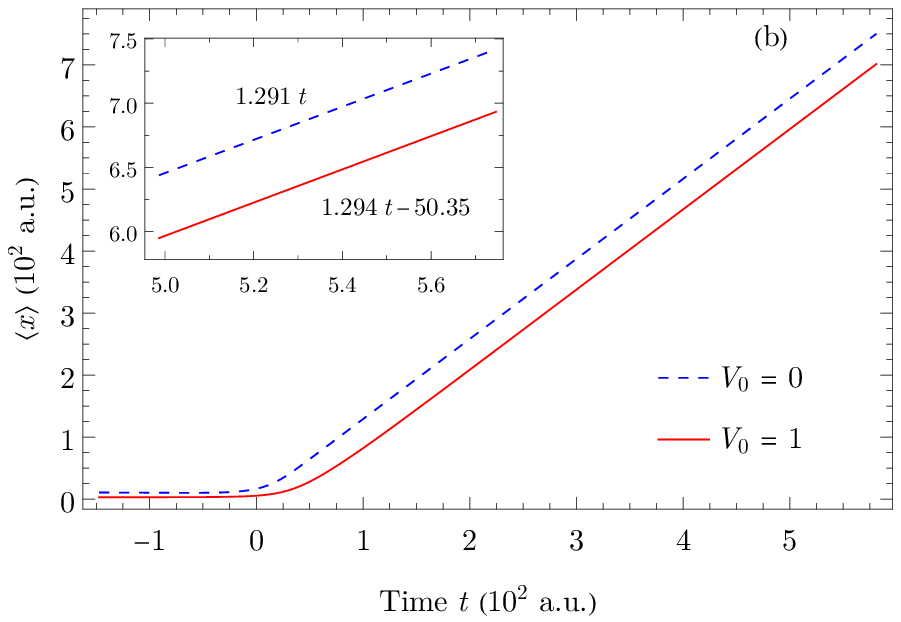}
  \caption{(Second Case) (a): the squared norm of the wavepackets for $V_0=0$ (blue dashed curve) and $V_0 \neq 0$ (red solid curve) are plotted versus distance at a large positive time. The two wavepackets have been rescaled to have the same normalization. The inset exhibits the Gaussian envelope $g(E)^2$, arbitrarily scaled, (blue dashed curve) over-plotted with the odd phaseshift energy derivative $\delta'_o(E)$ (red solid curve). (b): $\langle x \rangle$ for $V_0=0$ (blue dashed curve) and $V_0 \neq 0$ (red solid curve) is plotted versus time. The inset focuses on the asymptotics and shows their straight line fit equations. The parameters values are: $a=4,\ b=6,\ E_0=0.83695$,\ $s=20$, and $\alpha=1$.}
  \label{case2}
\end{figure}
The second case, presented in Fig.~\ref{case2}, treats a bandwidth comparable to the resonance width as demonstrated in the inset of Fig.~\ref{case2} (a). It is observed in (a) that the non-free wavepacket has acquired some skewness making it difficult to extract time delay information from peaks positions. However, the $\langle x \rangle$ curve asymptote (Fig.~\ref{case2} (b) inset)  still follows a straight line with coefficients agreeing very well with the ones calculated from convolution (Eqs.~(\ref{result2})). Note that in these first two cases, $E_0$ has been chosen to coincide with a resonant energy in the odd phaseshift such that the Gaussian $e^{-s^2(E-E_0)^2}$ is centered around the peak in the phaseshift derivative.

\begin{figure}[ht!]
  \centering
    \includegraphics[width=3.375in]{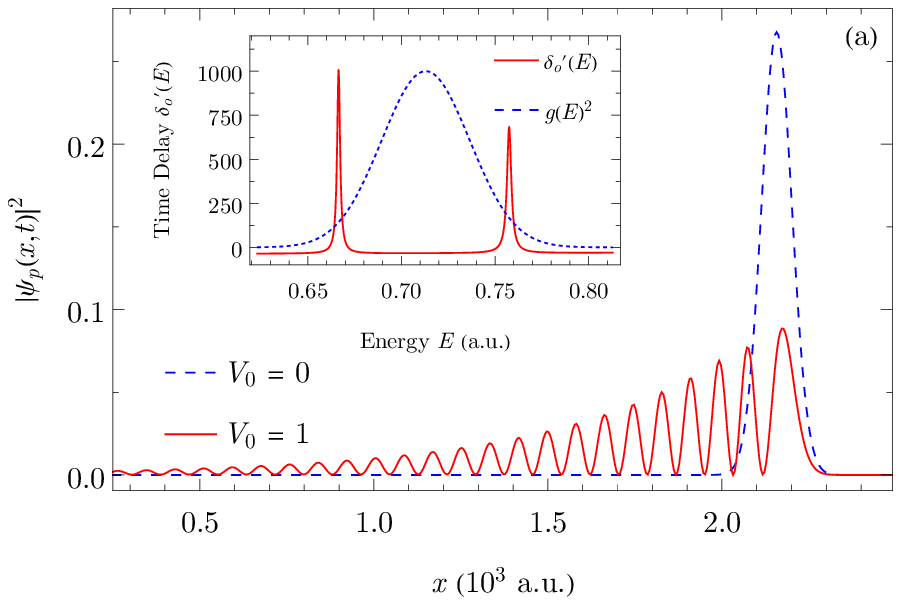}
    \\ \vspace{12pt}
    \includegraphics[width=3.375in]{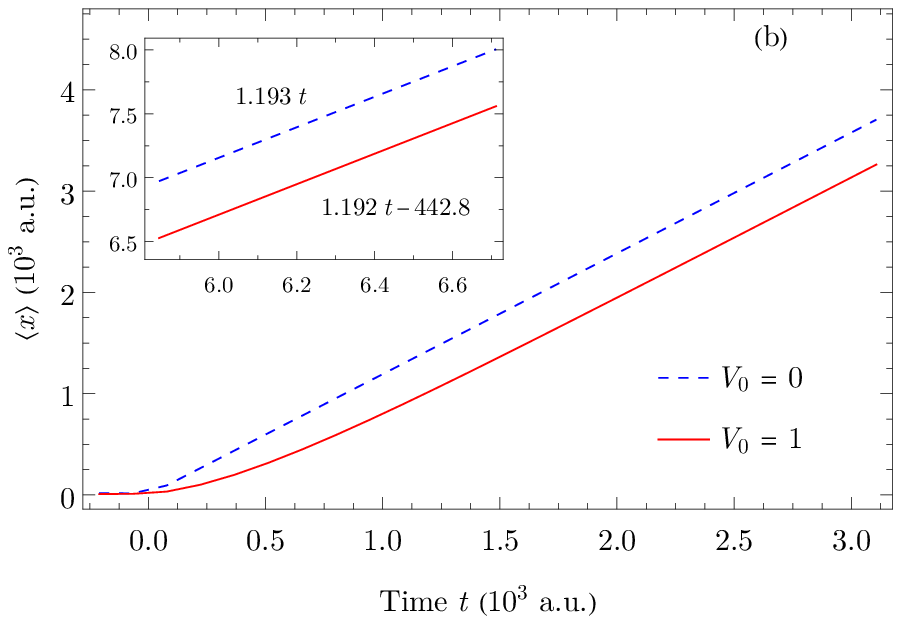}
  \caption{(Third Case) (a): the wavepackets norm squared for $V_0=0$ (blue dashed curve) and $V_0 \neq 0$ (red solid curve) are plotted versus distance at a large positive time. The two wavepackets have been rescaled to have the same normalization. The inset exhibits the Gaussian envelope $g(E)^2$, arbitrarily scaled, (blue dashed curve) over-plotted with the odd phaseshift energy derivative $\delta'_o(E)$ (red solid curve). (b): $\langle x \rangle$ for $V_0=0$ (blue dashed curve) and $V_0 \neq 0$ (red solid curve) is plotted versus time. The inset focuses on the curves asymptotes and shows their straight line fit equations. The parameters values are: $a=40,\ b=42,\ E_0=0.71298$,\ $s=30$, and $\alpha=1$.}
  \label{case3}
\end{figure}
Lastly, in the third case (Fig.~\ref{case3}), the potential well was made wider in order to add more resonances overlapping with the Gaussian envelope. As depicted in the Fig.~\ref{case3} (a) inset, values of $E_0$ and $s$ are set to excite two resonant states of the odd phaseshift. The main observation in this case is the non-free wavepacket (Fig.~\ref{case3} (a)) having several peaks as opposed to the first and second cases, which is due to multiple reflections occurring between the right barrier at $a<x<b$ and the origin. This claim is supported by noticing that the distance between each two consecutive peaks is approximately $2a = 80$ a.u. which is the additional distance traveled by each successive reflected wave. Consequently, the non-free wavepacket takes a relatively large time to completely tunnel outside the potential well since it has a long trail of decaying peaks. This can be deduced from the non-free $\langle x \rangle$ curve (Fig.~\ref{case3} (b)) which somehow struggles to immediately follow a straight line. To elaborate, note that in the first case the non-free wavepacket takes around $3s$ (in time) to fully tunnel outside the potential and it takes around $5s$ in the second case.  Conversely, in the third case, it takes the non-free wavepacket at least $50s$ to almost completely tunnel outside. Nonetheless, the $\langle x \rangle$ curve asymptote still follows, to good accuracy, the result in Eqs.~(\ref{result2}). The results of this case are analogous to the ones in Ref.~\onlinecite{Francis} involving ionizing Rydberg wavepackets.

\section{Conclusion} \label{conclusion}
In this article, we have analyzed a simple one-dimensional system with a dipole interaction to demonstrate that the liberated particle wavepacket admits a form of time delay relative to the free wavepacket not tunneling through the double barrier. This is manifested in Eq.~(\ref{result1}), in which a time delay, given directly by the phaseshift derivative $\delta'(E)$, appears in the analytic form of the wavepacket calculated for a narrow energy bandwidth. Moreover, scrutinizing the motion of $\langle x \rangle$ reveals that its large time asymptote incorporates a time delay given by convolution of the values of $\delta'(E)$ as shown in Eqs.~(\ref{result2}). In both results, the time delay source is the dipole matrix element since $\delta'_o(E) =\frac{d}{dE} \arg[{\bra{\psi_E^R}\hat{d}\ket{\psi_g}}]$. It is extremely essential to emphasize that the phaseshift $\delta$ entering in the time delay for photodetachment here is the same exact $\delta$ defined in scattering theory (i.e. the scattering phaseshift) as can be verified from Eqs.~(\ref{parity}). This is not a coincidence, as the same parameter $\delta$ enters in both the outgoing solution used for scattering and in the ingoing solution used for photoionization (see Section 4.3 of Ref.~\onlinecite{FanoRau}). This observation fits well with the perspective that all features of a quantum system, whether scattering phenomena involving continuum states or spectroscopic aspects involving bound states, can be described through the same parameters: the phaseshifts and their energy dependence, thus providing a coherent and consistent framework to work with. 

Note that our model system has no long range Coulomb potential, in contrast to the situation with true atomic and molecular photoionization. The Coulomb phaseshift is known to also be part of the full time delay in photoionization of neutral atoms or positive ions. However, this is usually a smooth contribution, so our conclusions about time delay near resonances are largely applicable to those Coulombic cases as well.

\acknowledgments
We thank A. R. P. Rau for numerous informative discussions. This work was supported in part by the U.S. Department of Energy,
Office of Science, Basic Energy Sciences, under Award
No. DE-SC0010545.

\appendix
\section{Equation \ref{Dirac} Derivation} \label{Appendix}
For the purposes of this derivation, we are going to make the approximation
\begin{equation}
    \psi_E^R(x) \xrightarrow[t \to \infty]{} \frac1{\sqrt{2 \pi k} } e^{i k x},
    \label{psiRapprox}
\end{equation}
which is not too crude. After all, this is the part of $\psi_E^R(x)$ that, upon being put in the wavepacket, survives in the large positive time limit as preliminarily demonstrated in Section~\ref{calcA}. Consider the quantity:
\begin{align}
    (E-E')\bra{\psi_{E'}^R}x\ket{\psi_E^R} &=\bra{\psi_{E'}^R}[x,\hat{H}_0]\ket{\psi_E^R} \nonumber\\&=\bra{\psi_{E'}^R}\frac{d}{dx}\ket{\psi_E^R}.
    \label{appeq1}
\end{align}
Equation~(\ref{psiRapprox}) allows us to write
\begin{align}
    \bra{\psi_{E'}^R}\frac{d}{dx}\ket{\psi_E^R} &= \frac{i}{2\pi} \sqrt{\frac{k}{k'}} \int e^{i (k-k') x} \,dx \nonumber \\
    &= i\sqrt{\frac{k}{k'}} \delta(k - k') \nonumber \\
    &= i\sqrt{k k'} \delta(E - E').
    \label{appeq2}
\end{align}
Hence, from Eq.~(\ref{appeq1}) and \ref{appeq2}, 
\begin{align}
    \bra{\psi_{E'}^R}x\ket{\psi_E^R} &= i\sqrt{k k'} \frac{\delta(E - E')}{E-E'} \nonumber \\
    &= -i\sqrt{kk'} \frac{d}{dE}\big[\delta(E - E')\big],
\end{align}
where a property of the Dirac-delta function derivative ($x \delta'(x) = -\delta(x)$) is used in the last step.

\bibliography{Main}

\end{document}